\documentclass[10pt,letterpaper]{article}
\usepackage{opex3}
\usepackage{amsmath,amssymb}        %
\usepackage[mathscr]{euscript}      
\usepackage{dsfont}                 
\usepackage{graphicx}
\usepackage[Symbolsmallscale]{upgreek} 
\usepackage[upmu]{gensymb}

\usepackage{color} 

\begin{document}

\title{Michelson interferometer with diffractively-coupled arm resonators in second-order Littrow configuration}

\author{Michael Britzger$^1$, Maximilian H. Wimmer$^1$, Alexander Khalaidovski$^1$, Daniel Friedrich$^1$,  Stefanie Kroker$^2$, Frank Br\"uckner$^2$, Ernst-Bernhard Kley$^2$, Andreas T\"unnermann$^2$, Karsten Danzmann$^1$, and Roman Schnabel$^{1,*}$}
\address{$^1$Albert-Einstein-Institut, Max-Planck-Institut f\"ur Gravitationsphysik and Leibniz Universit\"at Hannover, Callinstr. 38, 30167 Hannover, Germany\\$^2$Institut f\"ur Angewandte Physik, Friedrich-Schiller-Universit\"at Jena, Max-Wien-Platz 1, 07743 Jena, Germany\\}
\email{$^*$roman.schnabel@aei.mpg.de}


\begin{abstract}
Michelson-type laser-interferometric gravitational-wave (GW) observatories 
employ very high light powers as well as transmissively-coupled Fabry-Perot arm resonators in order to realize high measurement sensitivities. Due to the absorption in the transmissive optics, high powers lead to thermal lensing and hence to thermal distortions of the laser beam profile, which sets a limit on the maximal light power employable in GW observatories. Here, we propose and realize a Michelson-type laser interferometer with arm resonators whose coupling components are all-reflective second-order Littrow gratings. In principle such gratings allow high finesse values of the resonators but avoid bulk transmission of the laser light and thus the corresponding thermal beam distortion. The gratings used have three diffraction orders, which leads to the creation of a second signal port. We theoretically analyze the signal response of the proposed topology and show that it is equivalent to a conventional Michelson-type interferometer. In our proof-of-principle experiment we generated phase-modulation signals inside the arm resonators and detected them simultaneously at the two signal ports. The sum signal was shown to be equivalent to a single-output-port Michelson interferometer with transmissively-coupled arm cavities, taking into account optical loss. The proposed and demonstrated topology is a possible approach for future all-reflective GW observatory designs.
\end{abstract}



\section{Introduction}
Today's laser-interferometric gravitational-wave (GW) detectors~\cite{A09, A08, G10}, as well as the proposals for future observatories~\cite{ET11}, are based on advanced Michelson-type interferometer topologies. A gravitational wave will cause a differential change of the interferometer's arm lengths and will thus result in a signal at the detection output port. If the measurement sensitivity is limited by quantum shot noise, the signal-to-noise ratio (SNR) scales with the square root of the laser light power in the interferometer's arms. 
Issues that will effectively limit the maximal laser power employable in future GW observatories arise from thermal effects at mirror surfaces as well as inside the interferometer optics used in transmission. These effects are caused by the residual absorption within the surface and the bulk material, respectively, leading to a locally varying temperature increase. Because the index of refraction is a function of temperature, the absorption manifests itself as thermal lensing and thus as a reduced interferometer contrast \cite{lensing}. Furthermore, transmission through optical components generally leads to thermo-refractive noise and photo-thermo-refractive noise \cite{BGV00,Levin98}. 

\begin{figure}[b]
\centerline{\includegraphics[width=13cm]{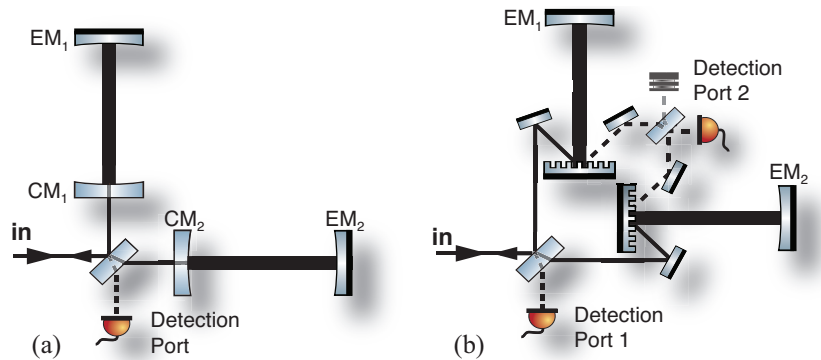}}\label{Fig:IFO}
 \caption{(a) Michelson-type laser interferometer with conventional (transmissively-coupled) Fabry-Perot arm cavities, consisting of the coupling mirrors CM$_\mathrm{1,2}$ and the highly reflective end mirrors EM$_\mathrm{1,2}$. (b) Interferometer with diffractively-coupled arm cavities. Because of the second-order Littrow configuration, the signal is distributed into two ports that both need to be monitored to obtain the full signal strength.}
\end{figure}

A promising approach to avoid thermal effects associated with absorption in the bulk material is the employment of an all-reflective topology. For this, diffraction gratings have been proposed as a means to split and recombine monochromatic light without transmission through a beam splitter or a cavity-coupling mirror~\cite{Drever}. An additional benefit is that in an all-reflective interferometer highly transmissive materials are no longer essential. This permits the use of opaque or less transmissive but mechanically and thermally favourable materials that would allow cooling down to temperatures not appropriate with current materials~\cite{Nawrodt10}.

In \cite{Sun}, the diffractive replacement of the balanced beamsplitter in Michelson- and Sagnac-interferometers was reported. In these experiments holographic metal gratings with an optical loss of about $3.6$\,\% were used. Meanwhile, dielectric reflection gratings have become a promising alternative. Such gratings are etched either directly into a substrate or into a multilayer dielectric coating~\cite{clausnitzer05}. They have a lower optical loss and a higher damage threshold. Furthermore, the diffraction characteristics are more precisely controllable than in the case of traditional metal gratings owing to constantly improving design, electron beam writing and etching skills. An all-reflective Michelson interferometer with a dielectric diffractive beamsplitter was demonstrated by Friedrich~$et\,al.$ in~\cite{gratingbs}, where an optical loss of less than $0.2$\,\% was achieved.

To increase the circulating light power, the GW observatories LIGO and Virgo make use of transmissively-coupled Fabry-Perot cavities (FP cavities), as shown in the simplified sketch of Fig.~\ref{Fig:IFO}~(a). This means that -- besides the beam splitter -- the partially transmissive resonator coupling mirrors are also exposed to high thermal load, making all-reflective resonator couplers based on reflection gratings all the more interesting. An early approach was the use of high-efficiency gratings in first-order Littrow configuration \cite{Sun,higheff1}, having, however, the drawback of stringent restrictions on beam pointing and alignment~\cite{freisenoise}. An alternative with considerably relaxed requirements is provided by the so-called \emph{three-port} gratings used in second-order Littrow configuration. In this topology, the diffraction efficiency of the first diffraction order is used as the coupling efficiency to a resonator that is arranged perpendicular to the grating surface, while the angle of the incident laser beam and the grating period are chosen such that the second diffraction order is back-reflected towards the laser source. Consequently, to reach high cavity finesses, a low-efficiency coupling and thus very shallow grating structures are required. This considerably relaxes the demands on grating fabrication processes and suggests that cavities with well-defined Gaussian TEM${_{00}}$ modes are feasible. Low-loss three-port gratings have been investigated theoretically and experimentally \cite{bunkowski1,bunkowski3}. It was shown that the single-ended three-port-grating-coupled cavity can be employed as the arm cavity of a GW interferometer. On resonance, no carrier power is lost to the additional port because the light fields involved interfere destructively~\cite{bunkowski2}. Hence, all power is reflected to the beamsplitter of the interferometer, making an optimal power-recycling possible \cite{burmeister2010,britzger2011a}. Thus, second-order Littrow gratings are promising optical components for future high-power and all-reflective gravitational wave observatories beyond the 3rd observatory generation.

Fig.~\ref{Fig:IFO}~(b) shows a schematic of the proposed topology with the arm cavities coupled in second-order Littrow configuration. A theoretical discussion of the signal response of the diffractively-coupled interferometer is presented in section~\ref{sec.theory}. It is shown that the sum signal of the two detection ports carries the same amount of information as the signal of a single-output-port Michelson interferometer with FP cavities shown in Fig.~\ref{Fig:IFO}~(a). Thus, the advantages of an all-reflective topology can be used without the drawback of a potential signal loss. In section~\ref{sec.experiment}, the table-top prototype interferometer and the signals measured in the two detection ports are discussed.

\section{Signal transfer function of the second-order Littrow grating cavity}
\label{sec.theory}
Laser GW interferometers are operated close to their \emph{dark fringe}, which means that due to destructive interference almost no (carrier) light leaves the signal port and all optical power is reflected back to the laser. If a gravitational wave interacts with the light fields in the FP cavities, phase modulation sidebands are generated. In the case of a conventional (transmissively coupled) arm cavity, the \emph{full} signal interferes constructively at the interferometer's signal port, manifesting itself as an amplitude modulation on the residual carrier light. One single detection port is therefore sufficient to gather the full information available as illustrated in Fig.~\ref{Fig:IFO}~(a). 

\begin{figure}[h]
\centerline{\includegraphics[width=13cm]{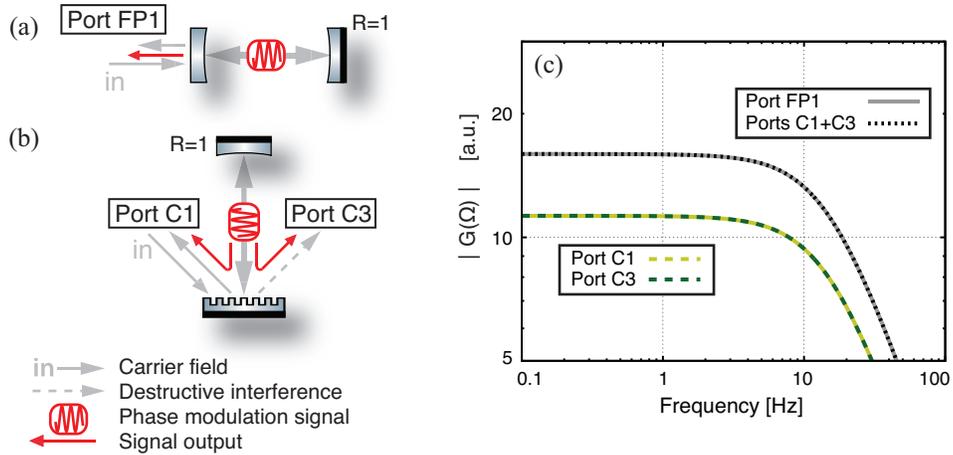}}\label{Fig:Trans}
 \caption{(a) Signal response of a single-ended standing-wave Fabry-Perot (FP) cavity. Both, signal and carrier light fields are back-reflected towards the laser source. (b) Signal response of a single-ended three-port grating cavity with reflection ports C1 and C3. While the carrier interferes destructively at C3, the signal is distributed equally among the two ports. (c) Phase-quadrature readout at the FP output port and at the ports C1/C3. The dotted line shows the sum of the C1- and C3-signals, being identical to the FP reference. As cavity parameters, values realisable in LIGO were chosen: $L=4$\,km, $\rho_0^2=1-2\eta_1^2=99.5\,\%$, $\rho_1^2=1-\tau_1^2=99.5\,\%$, $\eta_2=\eta_{2{\mathrm{min}}}$. The two cavities were tuned to resonance, the intra-cavity power was identical.}
\end{figure}

Fig.~\ref{Fig:Trans}~(a) shows a linear Fabry-Perot cavity with an end mirror reflectivity of $R=1$ (single-ended). Independently of the cavity detuning (length), the full carrier light power is back-reflected towards the source. Consequently, any phase modulation signal generated inside this cavity (e.\,g.~by a gravitational wave or equivalently by a moving end mirror or an electro-optical modulator) is fully coupled out at the retro-reflection port. The normalized frequency-dependent cavity-induced signal amplification for the upper and lower sidebands ($\pm\Omega$) reads
\begin{equation}
	\mathrm{g}_{\mathrm{FP1}}(\pm\Omega)=\frac{\mathrm{i}\tau_1 \exp[{\mathrm{i}(\Phi\pm{\Omega L}/{\mathrm{c}})}]} {1-\rho_1\rho_2 \exp[{2\mathrm{i}(\Phi\pm{\Omega L}/{\mathrm{c}})}]}\label{Eq:SF_FP}.
\end{equation}
Here, $\tau_{1}$ and $\rho_{1,2}$ are the amplitude transmissivities and reflectivities of the mirrors, $\Phi$ is the cavity detuning parameter, $\Omega$ the modulation frequency, $L$ the cavity length, and c the speed of light \cite{HeinzelDis}. When the cavity is tuned to resonance, $\Phi$ is equal to zero and the two sidebands are amplified by the same factor.

In the case of a diffractively-coupled single-ended arm cavity in second-order Littrow configuration that is illustrated in Fig.~\ref{Fig:Trans}~(b), the signal output is additionally influenced by the cavity detuning and the grating parameters~\cite{bunkowski1,bunkowski3}. If the second-order diffraction efficiency is minimal and the cavity is tuned to resonance, the carrier field is, as in the case of the linear Fabry-Perot cavity discussed above, fully back-reflected towards the source. This means that the interference for the carrier light is constructive at the input port (C1 in Figure~\ref{Fig:Trans}\,(b)) and correspondingly destructive at the forward-reflection port C3~\cite{bunkowski2}. The phase modulation signal generated inside the cavity, however, is split equally into the back-reflected port C1 and the forward-reflected port C3 because of the grating structure's symmetry~\cite{clausnitzer05}.

The signal amplification function is given by
\begin{equation}
	\mathrm{g}_{\mathrm{C1}}(\pm\Omega)=\mathrm{g}_{\mathrm{C3}}(\pm\Omega)=\frac{\eta_1 \exp[{\mathrm{i}(\phi_1+\Phi\pm{\Omega L}/{\mathrm{c}})}]}{1-\rho_0\rho_2 \exp[{2 \mathrm{i}(\Phi\pm{\Omega L}/{\mathrm{c}})}]},
\label{Eq:SF-grating}
\end{equation}
where $\eta_1$ is the diffraction efficiency and $\phi_1$ the phase shift associated with the first diffraction order, $\rho_0$ the amplitude reflectivity of the grating at normal incidence, and $\rho_2 = 1$ the amplitude reflectivity of the cavity end mirror \cite{bunkowski1}. If phase-modulation signal sidebands are generated in a single cavity that is tuned to resonance, an optimal readout (gathering the full information) has to be carried out in the phase quadrature. The signal transfer function for the phase quadrature readout can be written as
\begin{eqnarray}
	\mathbf{G}(\Omega)&=&   \mathrm{g}(+\Omega)-\mathrm{g}^*(-\Omega),\label{eq:G_approx_1}
\end{eqnarray}
assuming the normalized carrier to be real and positive~\cite{HeinzelDis}. Fig.~\ref{Fig:Trans}~(c) shows the phase quadrature readouts $\left|\mathbf{G}(\Omega)\right|$ in the \emph{single} signal port of a resonant Fabry-Perot cavity and in the \emph{two} signal ports of a three-port-grating coupled cavity with minimal second-order diffraction efficiency. For better comparison, the parameters were chosen such that the normal-incidence power reflectivity $\rho_0^2$ of the grating equals the power reflectivity $\rho_1^2$ of the Fabry-Perot cavity coupling mirror and the intra-cavity optical powers are identical. In this case, the signals of the grating-coupled cavity (dashed lines) are each a factor of two smaller than the FP cavity signal output (solid line). Thus, the sum signal (dotted line) is identical with the Fabry-Perot one, so that the two topologies are equivalent with respect to the signal-to-noise ratio. If, in contrast, merely the conventional output port is monitored (the detection port in Fig.~\ref{Fig:IFO}~(a), corresponding to detection port 1 of Fig.~\ref{Fig:IFO}~(b)), exactly 50\,\% of the signal is lost. Please note that the laser power input to a three-port grating cavity needs to be a factor of two higher in order to achieve the same power build-up as in a linear cavity \cite{bunkowski1}.

\section{Experimental setup and results}
\label{sec.experiment}
\begin{figure}[b]
\centerline{\includegraphics[width=13cm]{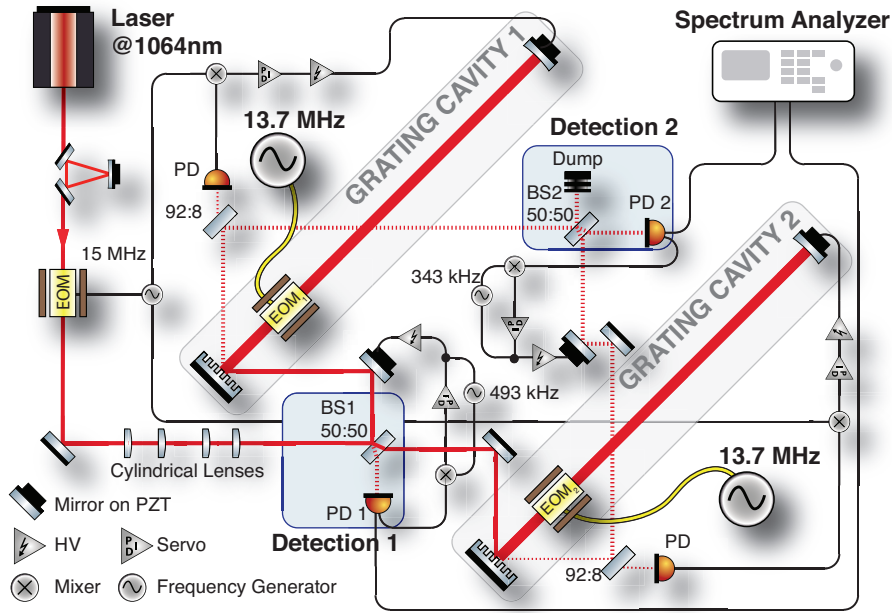}}\label{Fig:Setup}
 \caption{Schematic of the prototype experiment. The main interferometer was operated "close to the dark fringe" at the detection port, and thus almost all carrier light was back-reflected to the laser. To generate phase-modulation signals in the arm cavities, two electro-optical modulators (EOMs) were used. The EOMs were driven by phase-locked frequency generators, the signal frequency was 13.7\,MHz. The signals were brought to interference at the beam splitters BS1 and BS2 and recorded at the photo detectors PD1 and PD2, respectively. The photo detector output was analyzed with a R\&S\textsuperscript{\textregistered} FSP spectrum analyzer.}
\end{figure}

Figure~\ref{Fig:Setup} shows the layout of the experiment. The laser source was a single-mode Nd:YAG laser (non-planar ring oscillator, NPRO) operating at 1064\,nm. The laser output was transmitted through a ring mode-cleaner cavity to provide a spectrally and spatially filtered beam in the TEM${_{00}}$ mode \cite{PMC}. A set of cylindrical lenses was employed to mode-match this beam to the eigenmode of the grating arm cavities~\cite{britzger2011a}. 

The two diffractive cavity couplers used in the experiment were cut from a single dielectric three-port grating. The binary grating structure was written and etched in the topmost SiO$_2$-layer of a highly reflective multilayer coating applied onto a $1{\verb+"+}\times1{\verb+"+}$ large fused silica substrate. The grating had a period of $d=1450$\,nm for a first-order diffraction angle of $0^\circ$ and a second-order Littrow angle of incidence of $47.2^\circ$ at a laser wavelength of 1064\,nm \cite{clausnitzer05}. The grating design was chosen such that for a first-order diffraction efficiency of $\eta_1^2=3.3\,\%$ the second-order diffraction efficiency $\eta_2^2=0.04\,\%$ was close to the theoretical minimal boundary value~\cite{bunkowski1}. The grating was first characterized via a finesse measurement using the set-up discussed in \cite{bunkowski2} and then cut into two parts to provide two identical diffractive cavity couplers.

The arm cavities had a length of $L=81.5\,$cm each. To generate the phase-modulation signals simulating the effect of a GW, electro-optical modulators (EOMs) were placed in each arm, applying a modulation at 13.7\,MHz. The highly-reflective cavity end mirrors were mounted on piezoelectric transducers (PZTs). To stabilize the cavity length, a Pound-Drever-Hall (PDH) locking scheme~\cite{drever} was used. The modulation frequency for the PDH error signals was 15\,MHz. The error signals were detected at the respective forward-reflected grating port C3 using a partially transmissive mirror with a power transmission of 8\,$\%$. With the two cavities being resonant, the contrast at the main interferometer beam splitter BS1 was 98.7\,\%\,. The main interferometer was locked via an internal modulation scheme \cite{garching1988}. For this, phase-modulation sidebands at 493\,kHz were generated using the steering mirror located in the coupling path of grating cavity 1. The error signal detected at PD1 was demodulated at the modulation frequency and fed back to the steering mirror. The chosen operation point of the interferometer was close to the dark fringe so that a local oscillator beam for a self-homodyne readout scheme was available. The signal fields transmitted at the forward-reflection ports C3 of the two arms were brought to interference at another balanced beam splitter (BS2). To stabilize the phase relation of the two beam splitter input fields another internal modulation scheme at a sideband frequency of 343\,kHz was employed. The control signal was fed back to another PZT-mounted steering mirror as shown in Fig.~\ref{Fig:Setup}. At the second beamsplitter a contrast of 96.0\,\% was realized. The AC-gains of the two photo detectors PD1 and PD2 were carefully matched.

\begin{figure}[b]
\centerline{\includegraphics[width=13cm]{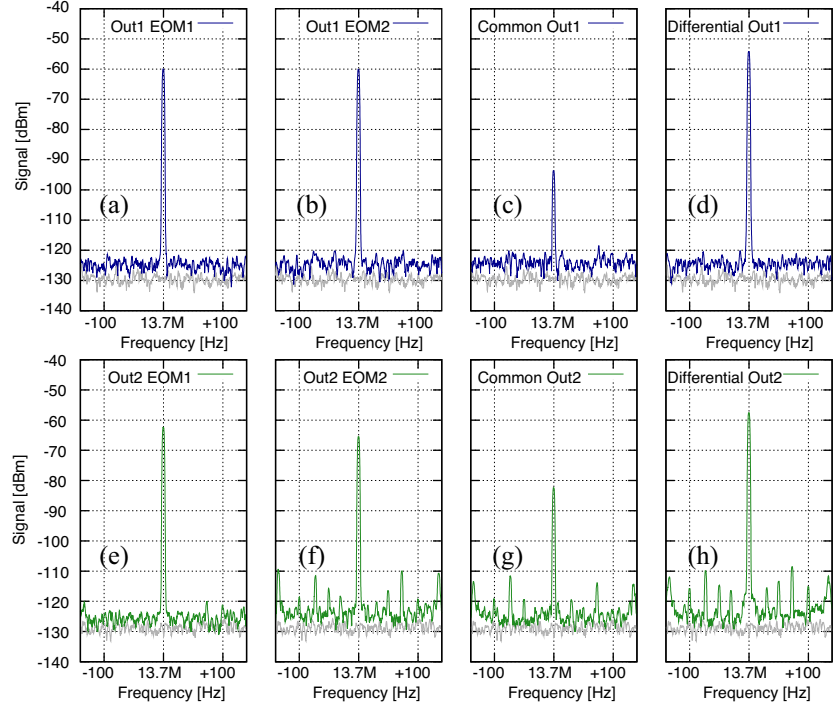}}
 \caption{(a,b) 13.7\,MHz signal generated by EOM$_{1,2}$ and detected by PD$_1$. In these measurements, no signal was generated simultaneously by the other EOM. The modulation depth was adjusted to generate $-$60\,dBm-signals. (c) EOM$_1$ and EOM$_2$ were operated simulteneously, the relative phase between the driving electric field was adjusted for maximally destructive interference. (d) same, the relative phase was shifted by $\pi$ to obtain maximally constructive interference. The signal amplitudes add up coherently, leading to a signal power increase by 6\,dB. (e-h) same as (a-d), but detected by PD$_2$. For a discussion of signal strengths and loss see text.}\label{Fig:Data}
\end{figure}

Figure~\ref{Fig:Data} shows the results obtained at the two signal output ports. Trace (a) shows the signal that was measured by PD1 if only EOM$_1$ in the north arm was actuated. The modulation depth was adjusted to generate a signal with a peak power of -60\,dBm for a resolution bandwidth of 3\,Hz. An equally strong signal was produced in grating cavity 2 using EOM$_2$ as shown in Fig.~\ref{Fig:Data}\,(b). If the EOMs were now actuated simultaneously, the phase relation of the EOM's signal generators determined the interference of the signals leaving the two arm cavities and combined at BS1. The maximal destructive interference, leading to a residual signal of $-$94\,dBm, was realized at a phase relation of 63$^{\circ}$ and is shown in trace (c). The deviation from the theoretical value of 0$^{\circ}$ is due to the residual differences of the two gratings (e.\,g.~originating from the cutting process and local variations), and to different signal responses of the two EOMs as well as to differing electrical signal paths (cabling).

When a phase shift of 180$^{\circ}$ was applied to one of the EOMs, the maximal constructive interference of the two signals was achieved, the results are shown in Figure~\ref{Fig:Data}\,(d). As expected, the signal increased by 6\,dB or a factor of four to $-$54\,dBm. The factor of \emph{four} is due to the fact that the two signal \emph{amplitudes} add up coherently when the two EOMs are operated in differential mode. The signal power that is proportional to the square of the sum amplitude thus increases by a factor of $2^2$.

The lower part of Figure~\ref{Fig:Data} shows the signals measured in detection port 2. The signals generated by EOM$_1$ and EOM$_2$ were $-$62.4\,dBm and $-$65.4\,dBm, respectively~[Fig.~\ref{Fig:Data}\,(e,f)]. The signal loss with respect to port 1 was due to a combination of propagation loss and unequal electronic stabilization loops. For EOM$_1$ the total signal loss was 43\,\%, and 71\,\% for EOM$_2$. The optical propagation loss was due to the partially transmissive steering mirror required for error signal generation as well as to absorption by optical components. Furthermore, the optical path to port 2 considerably exceeded the path length to port 1, so that air perturbations had a stronger effect in terms of beam pointing fluctuations and thus manifested itself as a fluctuating fringe visibility. The stronger loss for EOM$_2$ was mainly due to the PZT-mounted steering mirror in this path that led to beam pointing fluctuations. In addition, the length stabilization loop of grating cavity 2 had a lower stability than the one for grating cavity 1. 

The constructive interference of the two EOM signals in port 2 is shown in Figure~\ref{Fig:Data}\,(h). The signal strength was $-$57.5\,dBm and thus corresponded the measurements of the single EOM signal levels~[Fig.~\ref{Fig:Data}\,(e,f)]. For all measurements, the shot noise level was with a value of about $-$125\,dBm similar in the two ports, as was the dark noise with $-$129.5\,dBm. The signal-to-noise ratio was limited by quantum shot noise, no technical laser noise was present at the frequencies of interest. The sum of the signals recorded in port 1 and port 2 [Figure~\ref{Fig:Data}\,(d) and (h)] is by 26\,\% smaller than the value theoretically expected for the topology in the case that no signal loss is present. It is, however, still by 46\,\% larger than the signal from the single port 1. This experimentally confirms the theoretical concept of the proposed topology and its property of having the same measurement sensitivity as a single-output-port Michelson interferometer with transmissively-coupled FP arm cavities.

\section{Conclusion} 
We have proposed and analyzed a Michelson-type laser interferometer with diffractively-coupled arm resonators. A proof-of-concept table-top experiment was set up using dielectric binary-structured three-port gratings with minimal second-order diffraction efficiency in a second-order Littrow configuration. This topology introduces a second signal output port in addition to the one in a conventional Michelson-type interferometer. The signal generated inside the arm cavities splits equally into the two ports. We have theoretically shown that the full signal power can still be recovered if the two signal ports are monitored and that this signal power is identical to the one of a single-output-port Michelson interferometer in the case of equal cavity parameters and intra-cavity powers. This result was verified in a proof-of-principle experiment. The sum signal power was only slightly degraded by optical loss and imperfect electronic control loops. Our topology has an application in future precision metrology in all cases when light absorption in bulk optical materials sets a limit for the achievable measurement sensitivity. The conventional beam splitter employed in the work presented here can in principle be replaced by a purely reflective grating beam splitter, as already demonstrated in \cite{gratingbs}. In particular, we consider our topology to have an application in future (ground-based) gravitational wave observatories.

\section*{Acknowledgements} This work has been supported by Deutsche Forschungsgemeinschaft within the Sonderforschungsbereich TR7 and the Excellence Cluster QUEST, the IMPRS on gravitational wave Astronomy and the ET design study (the European Commission, FP7, Grant Agreement 211743).


\begin{thebibliography}{99}

\bibitem{A09} The LIGO Scientific Collaboration, ``LIGO: the Laser Interferometer Gravitational-Wave Observatory," Rep. Prog. Phys. {\bf72}, 076901 (2009).

\bibitem{A08} The Virgo Collaboration, ``Virgo: a laser interferometer to detect gravitational waves," {J. Instrum.} \textbf{7}, P03012 (2012).

\bibitem{G10} H.~Grote (for the LIGO Scientific Collaboration), ``The GEO 600 status,"  {Class.~Quant.~Grav.} \textbf{27}, 084003 (2010).

\bibitem{ET11} ET Science Team, ``Einstein gravitational wave telescope. Conceptual design study'', Internal report ET-0106C-10 (2011).

\bibitem{lensing} K.~A. Strain, K.~Danzmann, J.~Mizuno, P.~G.~Nelson, A.~R\"udiger, R.~Schilling and W.~Winkler,  ``Thermal lensing in recycling interferometric gravitational-wave detectors," {Phys. Lett. A} {\bf 194}, 124-132 (1994).
	 
\bibitem{BGV00} V.~B.~Braginsky, M.~L.~Gorodetsky, and S.~P.~Vyatchanin, ``Thermo-refractive noise in gravitational-wave antennae," Phys. Lett. A \textbf{271}, 303-307 (2000).

\bibitem{Levin98} Y.~Levin, ``Internal thermal noise in the LIGO test masses: A direct approach," {Phys. Rev. D} {\bf 57}, 659-663 (1998).

\bibitem{Drever} R.~W.~P.~Drever, ``Concepts for Extending the Ultimate Sensitivity of Interferometric gravitational-wave Detectors Using Non-Transmissive Optics with Diffractive or Holographic Coupling," in \textit{Proceedings of the Seventh Marcel Grossman Meeting on General Relativity}, M.~Keiser and R.~T.~Jantzen (eds.), World Scientific, Singapore (1995). 


\bibitem{Nawrodt10} R.~Nawrodt, S.~Rowan, J.~Hough, M.~Punturo, F.~Ricci, J.-Y.~Vinet, ``Challenges in thermal noise for 3rd generation of gravitational wave detectors,'' Gen.~Relativ.~Gravit. {\bf 43}, 593-622, (2011).

\bibitem{Sun} K.-X.~Sun and R.~L.~Byer, ``All-reflective Michelson, Sagnac, and Fabry-Perot interferometers based on grating beam splitters," Opt. Lett.  {\bf 23}, 567-569 (1997). 

\bibitem{clausnitzer05}T.~Clausnitzer, E.-B.~Kley, A.~T\"unnermann, A.~Bunkowski, O.~Burmeister, R.~Schnabel, K.~Danzmann, S.~Gliech, and A.~Duparr\'e, ``Ultra low-loss low-efficiency diffraction gratings," Optics Express  \textbf{13}, 4370-4378 (2005).

\bibitem{gratingbs} D.~Friedrich, O.~Burmeister, A.~Bunkowski, T.~Clausnitzer, S.~Fahr, E.-B.~Kley, A.~T\"unnermann, K.~Danzmann, R.~Schnabel, ``Diffractive beam splitter characterization via a power-recycled interferometer," {Opt. Lett.} {\bf 33}, 101-103 (2008).

\bibitem{higheff1} A.~Bunkowski, O.~Burmeister, K.~Danzmann, R.~Schnabel, T.~Clausnitzer, E.-B.~Kley, and A.~T\"unnermann, ``Optical characterization of ultra-high diffraction efficiency gratings," {Appl. Opt.} {\bf45}, 5795-5799 (2006).

\bibitem{freisenoise}A.~Freise, A.~Bunkowski, and R.~Schnabel, ``Phase and alignment noise in grating interferometers," New J. Phys. {\bf9}, 433 (2007).
	
\bibitem{bunkowski1} A.~Bunkowski, O.~Burmeister, K.~Danzmann, R.~Schnabel, ``Input-output relations for a 3-port grating coupled Fabry-Perot cavity," {Opt. Lett.} {\bf 30}, 1183-1185 (2005).

\bibitem{bunkowski3}A.~Bunkowski, O.~Burmeister, K.~Danzmann, R.~Schnabel, T.~Clausnitzer, E.-B.~Kley, and A.~T\"unnermann, ``Demonstration of 3-port grating phase relations," Opt. Lett. {\bf31}, 2384-2386 (2006).

\bibitem{bunkowski2} A.~Bunkowski, O.~Burmeister, P.~Beyersdorf, K.~Danzmann, R.~Schnabel, T.~Clausnitzer, E.-B.~Kley, A.~T\"unnermann, ``Low-loss grating for coupling to a high-finesse cavity," {Opt. Lett.} {\bf 29}, 2342-2344 (2004).

\bibitem{burmeister2010} O.~Burmeister, M.~Britzger, A. ~Th\"uring, D.~Friedrich, F.~Br\"uckner, K.~Danzmann, and R.~Schnabel, ``All-reflective coupling of two optical cavities with 3-port diffraction gratings," Optics Express {\bf18}, 9119-9132 (2010).

\bibitem{britzger2011a} M.~Britzger, D.~Friedrich, S.~Kroker, F.~Br\"uckner, O.~Burmeister, E.~B.~Kley, A.~T\"unnermann K.~Danzmann, and R.~Schnabel, ``Diffractively coupled Fabry-Perot resonator with power-recycling," Optics Express {\bf 19}, 14964-14975 (2011).

\bibitem{HeinzelDis}G.~Heinzel, ``Advanced optical techniques for laser-interferometric gravitational-wave detectors," PhD Thesis, Hannover (1999).

\bibitem{PMC}B. Willke, N. Uehara, E. K. Gustafson, R. L. Byer, ``Spatial and temporal filtering of a 10-W Nd:YAG laser with a Fabry-Perot ring-cavity premode cleaner," Opt. Lett. \textbf{23}, 1704-1706 (1998).

\bibitem{drever} R.\,W.\,P. Drever, J.\,L. Hall, F.\,V. Kowalski, J. Hough, G.\,M. Ford, A.\,J. Munley, and H. Ward, ``Laser Phase and Frequency Stabilization using an Optical Resonator," Appl. Phys. B. {\bf 31}, 97-105 (1983).

\bibitem{garching1988} D.~Shoemaker, R.~Schilling, L.~Schnupp, W.~Winkler, K.~Maischberger, A.~R\"udiger, ``Noise behavior of the Garching 30-meter prototype gravitational-wave detector," Phys. Rev. D {\bf38}, 423-432 (1988).

\end{thebibliography}
\end{document}